\documentclass[superscriptaddress,runinaddress,showpacs,showkeys,aps,prd,floatfix,onecolumn]{revtex4}
\usepackage{amssymb,amsmath,epsfig}
\usepackage{amsmath,graphicx}
\usepackage{changes}
\usepackage{palatino}
\usepackage[colorlinks=true,
            linkcolor=blue,
            urlcolor=blue,
            citecolor=blue]{hyperref}
\begin{document}

\title{Weak Gravitational lensing by  Bocharova-Bronnikov-Melnikov-Bekenstein black holes using Gauss-bonnet theorem}

\author{Wajiha Javed}
\email{wajiha.javed@ue.edu.pk} 
\affiliation{Division of Science and Technology, University of Education, Township-Lahore, Pakistan}

\author{Muhammad Bilal Khadim}
\email{blaljutt723@gmail.com}
\affiliation{Division of Science and Technology, University of Education, Township-Lahore, Pakistan}

\author{Ali {\"O}vg{\"u}n}
\email{ali.ovgun@emu.edu.tr}

\affiliation{Physics Department, Arts and Sciences Faculty, Eastern Mediterranean
University, Famagusta, North Cyprus via Mersin 10, Turkey.}

\date{\today}
\begin{abstract}
In this article, we demonstrate the weak gravitational lensing in the
context of Bocharova-Bronnikov-Melnikov-Bekenstein (BBMB) black hole.
To this desire, we derive the deflection angle of light in a plasma medium by BBMB black hole
using the Gibbons and Werner method. First, we obtain the Gaussian
optical curvature and implement the Gauss-Bonnet theorem to investigate
the deflection angle for spherically symmetric spacetime of BBMB black hole.
Moreover, we also analyze the graphical behavior of deflection angle by BBMB black hole
in the presence of plasma medium.
\end{abstract}

\pacs{95.30.Sf, 98.62.Sb, 97.60.Lf}

\keywords{Weak gravitational lensing; Black hole; Deflection angle; Gauss-Bonnet theorem}
\maketitle

\section{Introduction}
The discovery of first neutron collision \cite{2} was an epochal which is the beginning of gravitational wave astronomy. After gravitational waves was discovered, a large number of the modified gravity theories face a lot of problems. On the other hand, one of the important testing method of gravity is gravitational lensing. The light is deflected in the presence of a massive body which is predicted by General
Relativity. This phenomena is called gravitational lensing. Its experimental conformation was first given by Cavendish, Soldner \cite{14} as well as Einstein in
their observation to identify the gravitational deflection of light. The gravitational lensing has been classified in the literature as a strong lensing, weak lensing and micro lensing \cite{Bartelmann:1999yn,Cunha:2015yba}.

Weak gravitational lensing is a powerful tool to measure masses of a variety of objects in the universe. The weak gravitational lensing provides a way to find the mass of astronomical objects without requiring about their composition or dynamical states. Weak lensing also investigate the cause of the accelerated expansion of the universe and also distinguish between modified gravity and dark energy. For instance, the
lens equation is obtained for the Schwarzschild black hole by  Ellis and Virbhadra \cite{22} and
more comprehensively discussed by Virbhadra \cite{23}.
Many authors investigate the gravitational lensing by other astrophysical objects such as naked singularities black holes (BH), wormholes and few other related
objects \cite{g1}-\cite{g23}.

Gibbons and Werner (GW) proposed a new method to calculate weak deflection angle using the Gauss-Bonnet theorem (GBT) \cite{GW}. They showed that this method provides that the deflection angle is a global property. Sakalli and Ovgun applied this novel method to the Rindler modified \cite{A4}
Schwarzschild black hole and obtein the weak deflection angle. The Gibbons-Werner method has been applied by different authors on black holes as well as wormholes \cite{A1}-\cite{A28}. The GBT method is based on the following equation \cite{GW}: \begin{equation}
\alpha=-\int\int_{D\infty}\mathcal{K}dS.\nonumber\\
\end{equation}
Here, $\mathcal{K}$ and $dS$ represents the Gaussian curvature and surface element respectively.
After that, Werner \cite{W} extended this method and obtained the weak deflection of light by Kerr
black hole by using the Nazims's method for Rander-Finsler metric. Recently, Gallo and Crisnejo (GC) \cite{A28} discussed the deflection of light in the plasma medium. Since then the work of weak gravitational lensing is continuously through the method of GW by using the GBT for different black holes and wormholes.

In this article, our main aim is to investigate weak deflection angle of BBMB black hole. For this purpose, this article is categories as follows; In section 2 we review the BBMB black hole.
In section 3, by using the GBT we compute the weak deflection angle of BBMB black hole in plasma medium.
In section 4, we conclude our results.

\section{Bocharova-Bronnikov-Melnikov-Bekenstein black hole}

The BBMB black hole in a static and spherically symmetric form is given as \cite{BBMB}
\begin{equation}
ds^{2}=-f(r)dt^{2}+\frac{dr^{2}}{f(r)}+{r}^{2}(d\theta^{2}+\sin^{2}\theta d\phi^{2}),
\end{equation}
where the metric function $f(r)$ is
\begin{equation}
f(r)=1-\frac{2m}{r}+\frac{m^2}{r^2},
\end{equation}
here $m$ is a mass of BH.
The optical space in equatorial plane ($\theta=\frac{\pi}{2}$ ) to get null geodesic ($ds=0$)
\begin{equation}
dt^{2}=\frac{dr^{2}}{f(r)^2}+\frac{{r}^{2}d\phi^{2}}{f(r)}. \label{metric}
\end{equation}

To investigate the
weak gravitational lensing by BBMB black hole in plasma medium, we calculate the refractive index $n(r)$:
\begin{equation}
n(r)=\sqrt{1-\frac{\omega_{e}^{2}}{\omega_{\infty}^{2}}f(r)},
\end{equation}
where $\omega_{e}$ and $\omega_{\infty}$ are electron plasma frequency and light frequency calculated by an observer at
infinity respectively, then the corresponding optical metric illustrated as
\begin{equation}
d\sigma^{2}=g^{opt}_{ij}dx^{i}dx^{j}=\frac{n^{2}(r)}{f(r)}\left(\frac{dr^{2}}{f(r)}+{r}^{2} d\phi^{2}\right).
\end{equation}

The Gaussian optical curvature that is evaluated as follows:
\begin{equation}
\mathcal{K}=\frac{RicciScalar}{2}.\\
\end{equation}

After simplifying, Gaussian optical curvature for BBMB black hole in leading order term  is calculated as follow:

\begin{eqnarray}
\mathcal{K}&\approx&-\frac{2m}{r^{3}}-\frac{3m\omega_e^2}{\omega_\infty^2 r^3}+\frac{6m^2}{r^4}+\frac{17m^2\omega_e^2}{r^4\omega_\infty^2}.\label{AH6}
\end{eqnarray}

\section{Deflection angle of BBMB black holes and Gauss-Bonnet theorem}
In this section, we drive the deflection angle of light by BBMB BH using the GBT.
By using GBT in the region $\mathcal{H}_{R}$ with $\partial H_{R}=\gamma_{\bar{g}} \cup C_{R}$, given as \cite{GW}
\begin{equation}
\int_{\mathcal{H}_{R}}\mathcal{K}dS+\oint_{\partial\mathcal{H}_{R}}kdt
+\sum_{t}\epsilon_i=2\pi\mathcal{X}(\mathcal{H}_{R}),
\end{equation}
where, $k$ represent the geodesic curvature, and $\mathcal{K}$ denotes
the Gaussian optical curvature. One can define $k$ as $\kappa=\tilde{g}\left(\nabla_{\dot{\gamma}} \dot{\gamma}, \ddot{\gamma}\right)$
in such a way that $\tilde{g}(\dot{\gamma}, \dot{\gamma})=1$, where $\ddot{\gamma}$ represent
the unit acceleration vector and $\epsilon_i$ denotes the exterior angle
at $ith$ vertex respectively. As $R\rightarrow\infty$, we obtain the jump angles equal to $\pi/2$, hence total jump angles are $\epsilon_{i}+\epsilon_{ii} \rightarrow \pi$. Here, $\mathcal{X}(\mathcal{H}_{R})=1$ is a Euler
characteristic number and $\mathcal{H}_{R}$ denotes the non-singular domain. Therefore, we obtain \cite{GW}
\begin{equation}
\int\int_{\mathcal{H}_{R}}\mathcal{K}dS+\oint_{\partial
\mathcal{H}_{R}}kdt=\pi.
\end{equation}

 At $0^{th}$ order weak field deflection limit of the light for the straight-line approximation is defined as $r(t)=b/\sin\phi$. But we use the the first-order light ray trajectory as follows \cite{Jusufi:2017drg} 
 
 \begin{equation} \frac{1}{r_p}=\frac{\sin (\phi)}{b}+\frac{1}{2} \frac{m(3+\cos (2 \phi))}{b^{2}}+\frac{1}{16} \frac{m^{2}(37 \sin (\phi)+30(\pi-2 \phi) \cos (\phi)-3 \sin (3 \phi))}{b} \end{equation}
 and GW shows that the equation of GBT reduces to simple form \cite{GW}, then GC analyses the deflection angle in the plasma medium  by using this equation \cite{A28}:
\begin{equation}
\alpha=-\int^{\pi}_{0}\int^{\infty}_{r_p}\mathcal{K}\sqrt{detg^{opt}}drd\phi,\label{AH7}
\end{equation}
here we put the leading term of equation Eq. \ref{AH6} into above equation Eq. \ref{AH7},
so the obtained deflection angle of the photon rays are moving in a medium of homogeneous plasma up to leading order term is computed as:
\begin{eqnarray}
\alpha&\simeq& \frac{4m}{b}+\frac{2m\omega_{e}^{2}}{b \omega_{\infty}^{2}}-\frac{3m^{2}\pi }{4b^{2}}.\label{result}
\end{eqnarray}
Hence, the effect of plasma can be removed if $\left(\omega_{e}=0\right)$,  or $\left(\beta=\omega_{e} / \omega_{\infty} \rightarrow 0\right) $, and it reduces to vacuum case, and it is agreement with \cite{Jusufi:2015laa,Sereno:2003nd} up to the second
order in $m$, if second term is $Q^2$ instead of $m^2$. The value of the photon frequency in plasma medium is $\omega_{e} / \omega_{\infty}= 6 \times 10^{-3}$ \cite{BisnovatyiKogan:2010ar}.

\section{Conclusion}

The present article is about the investigation of deflection
angle by BBMB black hole in plasma medium.
In this regard, we analysis the weak gravitational lensing by using
GBT and get the deflection angle of light for BBMB black hole.

\begin{figure}[h]
 \centering
\includegraphics[width=0.6\linewidth]{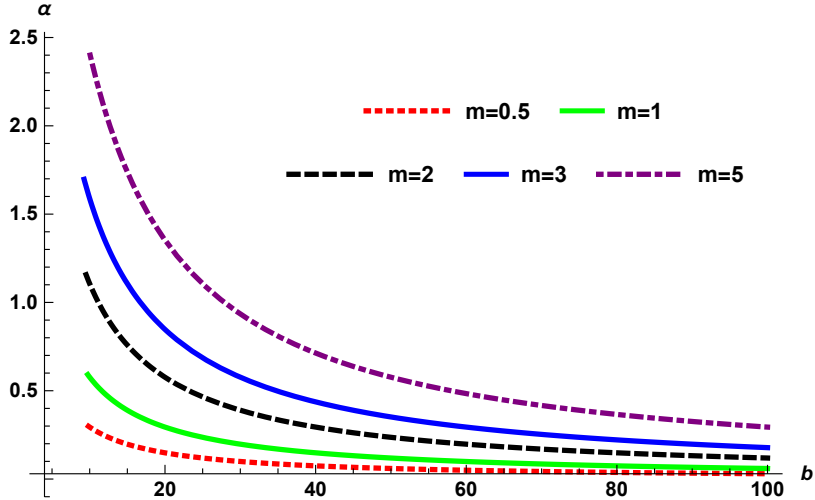}
\caption{ $\alpha$ versus $b$ by fixing the value of $\beta=1$ and varying $m$.}

\end{figure}

\begin{itemize}
\item  \textbf{Figure 1} indicates the behavior of deflection angle
w.r.t $b$ by fixing the value of $\beta$ and varying $m$.
It is to be observed that for values of increasing $m$, the behavior of deflection angle gradually increasing.
\end{itemize}

\begin{figure}[h]
 \centering
\includegraphics[width=0.6\linewidth]{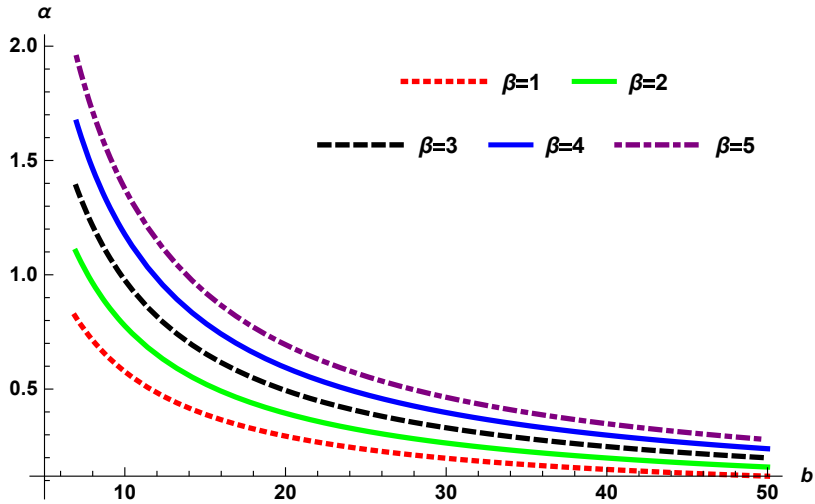}
\caption{ $\alpha$ versus $b$ by fixing the value of $m$ and varying $\beta$.}
\end{figure}

\begin{itemize}
\item  \textbf{Figure 2} represents the behavior of deflection angle w.r.t
$b$ by varying the mass $\beta$ and taking $m$ fixed.
We noticed that for values of $\beta>0$ the behavior of deflection angle gradually
increasing, on the other hand for the values of $\beta<0$
the behavior of deflection angle decreasing. 
\end{itemize}
The obtained deflection angle is given as follows:
\begin{eqnarray}
\alpha&\simeq& \frac{4m}{b}+\frac{2m\omega_{e}^{2}}{b \omega_{\infty}^{2}}-\frac{3m^{2}\pi}{4b^{2}}.
\end{eqnarray}

We examine that by the reduction of some parameters the obtained deflection angle converted
into the Schwarzschild deflection angle up to the first order terms. In addition we also discuss the graphical effect
of different parameters on deflection angle by BBMB in a plasma medium.

\end{document}